\documentclass[sigconf]{acmart}
\AtBeginDocument{%
  }

\usepackage{lipsum}
\usepackage{algorithm}
\usepackage{algpseudocode}
\usepackage{amsmath}
\usepackage{subcaption}
\usepackage{soul} 
\usepackage{multirow}
\usepackage{booktabs}
\usepackage{mathtools}

\copyrightyear{2025}
\acmYear{2025}
\setcopyright{cc}
\setcctype{by}
\acmConference[WWW Companion '25]{Companion Proceedings of the ACM Web Conference 2025}{April 28-May 2, 2025}{Sydney, NSW, Australia}
\acmBooktitle{Companion Proceedings of the ACM Web Conference 2025 (WWW Companion '25), April 28-May 2, 2025, Sydney, NSW, Australia}
\acmDOI{10.1145/3701716.3715486}
\acmISBN{979-8-4007-1331-6/25/04}

\begin{document}

% \title{The Name of the Title Is Hope}
% \title{Joining Collaborative Knowledge and Large Language Models for Conversational Recommendation}
% \title{Bridging Conversational Context and Collaborative Filtering Datasets for Conversational Recommendation}
\title{Bridging Conversational and Collaborative Signals for Conversational Recommendation}

\author{Ahmad Bin Rabiah}
\orcid{0009-0002-5991-4299}
\affiliation{
  \institution{University of California, San Diego}
  \city{La Jolla}
  \state{CA}
  \country{USA}
}
\email{abinrabiah@ucsd.edu}

\author{Nafis Sadeq}
\affiliation{
  \institution{University of California, San Diego}
  \city{La Jolla}
  \state{CA}
  \country{USA}
  }
\email{nsadeq@ucsd.edu}

\author{Julian McAuley}
\orcid{0000-0003-0955-7588}
\affiliation{
  \institution{University of California, San Diego}
  \city{La Jolla}
  \state{CA}
  \country{USA}
  }
\email{jmcauley@eng.ucsd.edu}

\renewcommand{\shortauthors}{Ahmad Bin Rabiah, Nafis Sadeq, and Julian McAuley}


\begin{abstract} 
Conversational recommendation systems (CRS) leverage contextual information from conversations to generate recommendations but often struggle due to a lack of collaborative filtering (CF) signals, which capture user-item interaction patterns essential for accurate recommendations.
We introduce Reddit-ML32M, a dataset that links Reddit conversations with interactions on MovieLens 32M, to enrich item representations by leveraging collaborative knowledge and addressing interaction sparsity in conversational datasets.
We propose an LLM-based framework that uses Reddit-ML32M to align LLM-generated recommendations with CF embeddings, refining rankings for better performance.
We evaluate our framework against three sets of baselines: CF-based recommenders using only interactions from CRS tasks, traditional CRS models, and LLM-based methods relying on conversational context without item representations. Our approach achieves consistent improvements, including a 12.32\% increase in Hit Rate and a 9.9\% improvement in NDCG, outperforming the best-performing baseline that relies on conversational context but lacks collaborative item representations.
\end{abstract}


% \keywords{Do, Not, Us, This, Code, Put, the, Correct, Terms, for,
  % Your, Paper}

\newcommand{\gpt}{\texttt{GPT-3.5-t}}
\newcommand{\sellname}{BridgeCRS}


\maketitle

\section{Introduction}

\begin{figure*}
    \centering
    \includegraphics[width=.98\linewidth,keepaspectratio]{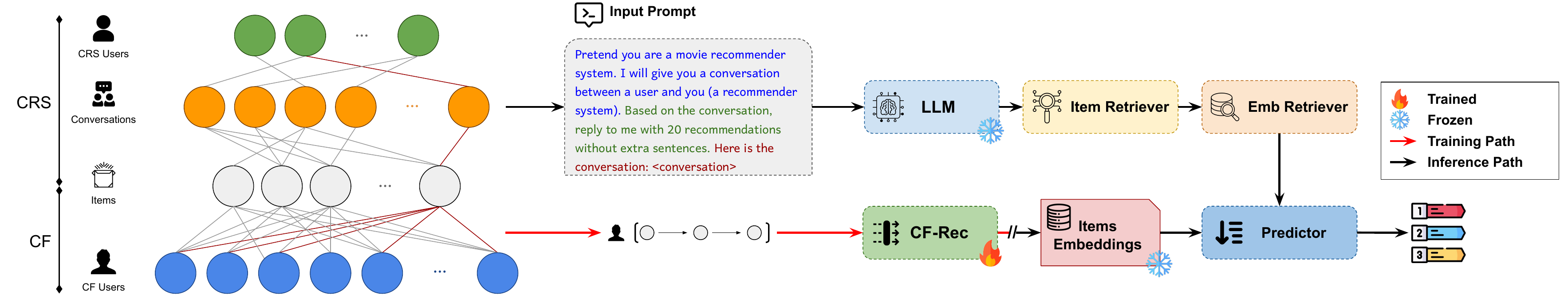}
    \caption{Illustration of the proposed CRS framework, which integrates LLM-based conversational recommendations with CF signals. The predictor uses these signals to generate ranked recommendations.}
    \label{fig:teaser}
\end{figure*}

Conversational recommender systems (CRS) aim to elicit user preferences and generate recommendations through interactive dialogues~\cite{he2023large, li2018towards,sun2018conversational,wang-etal-2023-rethinking-evaluation}. Unlike traditional recommenders, CRS leverages conversational context to adapt to user preferences. However, existing CRS approaches face two main challenges:
(1) They do not leverage collaborative filtering (CF) signals with conversational context, leading to suboptimal recommendations~\cite{zhou2020improving, friedman2023leveraging}, and 
(2) CRS datasets are typically sparse, making it difficult to develop robust item representations for accurate recommendations.
To this end, a typical CRS consists of two components: (1) a \textit{generator} to produce natural language responses and (2) a \textit{recommender} to rank items aligned with user preferences~\cite{he2023large, li2018towards, chen2019towards}.

\textbf{From a CRS model perspective}, CF-based approaches excel at modeling user-item relationships but fail to leverage conversational context~\cite{salem2014history, kang2018self}. 
Similarly, knowledge-driven methods improve semantic understanding using external knowledge graphs but lack the collaborative signals needed to model item relationships~\cite{zhou2020improving}.
Large language models (LLMs) show promise in CRS due to their natural language understanding capabilities~\cite{brown2020language,chowdhery2023palm,ouyang2022training}, but they rely at most on textual metadata, which constrains their ability to capture item relationships~\cite{he2024reindex, yang2024item}.
Recent methods integrate CF embeddings with LLMs to improve recommendation tasks. For instance, A-LLMRec incorporates CF embeddings into prompts, while CoLLM maps them onto token embeddings~\cite{kim2024large, zhang2023collm}. 
ILM aligns CF embeddings with textual embeddings but lacks conversational context~\cite{yang2024item}. 
While these methods incorporate CF with LLMs to improve recommendations, they are not designed to utilize conversational context. Our framework bridges this gap by integrating CF signals and conversational data into a unified CRS approach.

\textbf{From CRS dataset perspective}, current CRS datasets suffer from low interactions per item. For example, ReDIAL dataset~\cite{li2018towards} contains 31,988 interactions, with 78.98\% of items having fewer than 10 interactions~\cite{he2024reindex}. Similarly, Reddit-Movie~\cite{he2023large} has 51,148 for 31,396 items as shown in Table~\ref{tab:dataset_stats}. 
To our knowledge, no existing dataset bridges conversational context with CF information.
To address these limitations, we propose a dataset that integrates CF signals with conversations and mitigates sparsity in CRS datasets.

In this work, we make two key contributions: 
\textbf{First}, we construct a new \textbf{Reddit-ML32M} dataset by linking the Reddit-Movie dataset~\cite{he2023large} with the MovieLens 32M dataset~\cite{harper2015movielens}. This dataset enriches the interaction space, increasing the total number of interactions from 51,148 to over 30 million and improving dataset density from 0.013\% to 0.68\%, as shown in Table~\ref{tab:dataset_stats}. The dataset facilitates the development of richer item representations for CRS tasks by exploiting both conversational data with CF information at the item level.
\textbf{Second}, we propose a framework that integrates conversational context from LLMs with CF-based item representations as depicted in Figure~\ref{fig:teaser}. GPT-3.5-turbo~\cite{gpt-models}\footnote{We refer to this model as GPT-3.5t hereafter} is used in a zero-shot setting as a backbone to generate initial recommendations from conversational context, which are then refined by ranking them against item embeddings generated by a pre-trained SASRec model~\cite{kang2018self}. This approach mitigates biases in LLM outputs, such as over-reliance on popular items or training data~\cite{he2023large, yang2024item}, while incorporating collaborative signals.
Empirical evaluation across three paradigms: CF-based recommenders, CRS models, and LLM-based CRS without CF signals shows the effectiveness of our framework. Our approach achieves a 12.32\% improvement in Hit Rate and a 9.9\% improvement in NDCG over the best-performing baseline GPT-3.5t~\cite{he2024reindex}.

\begin{table}[t]
    \centering
    \caption{Statistics for Reddit-Movie~\cite{he2023large} and Reddit-ML32M datasets to improve item representations for CRS tasks.}
    \label{tab:dataset_stats}
    \vspace{-10pt}
    {\small
    \begin{tabular}{@{}lrrrrr@{}} 
        \toprule
        \textbf{Dataset} & \textbf{\#Interactions} & \textbf{\#Users} & \textbf{\#Items} & \textbf{Density}  \\ 
        \midrule
        Reddit    & 51,148 & 12,508 & 31,396 & 0.013\%          \\
        Reddit-ML32M    & 30,074,259 & 200,947 & 22,014 & 0.68\%          \\
        \bottomrule
    \end{tabular}}
\end{table}
\section{Method}

\subsection{Problem Formulation}
Let $\mathit{U} \coloneqq \{u_1, \dots, u_M\}$ and $\mathit{I} \coloneqq \{v_1, \dots, v_N\}$ represent sets of $M$ users and $N$ items, respectively. We define $\mathit{W}$ as the vocabulary of words forming an utterance $s \coloneqq (w_i)_{i=1}^m$, with each $w_i \in \mathit{W}$. 

\subsubsection*{\textbf{Conversation Modality}} 
We model a multi-turn conversation for recommendation tasks as $C \coloneqq (u^t, s^t, \mathit{I}^t)_{t=1}^T$ with $T$ turns, where $\mathit{I}^t \subseteq \mathit{I}$ represents the items mentioned in turn $t$. For each turn, $s^t$ is the utterance, $u^t$ is the user in the conversation, and $\mathit{I}^t$ may be empty if no items are mentioned. A user seeking recommendations is a \textit{seeker}, while a \textit{recommender} provides recommendations.

\subsubsection*{\textbf{Interaction Modality}} 
Observed user-item interactions are defined as $\mathit{R} \coloneqq \{(u, v) \mid u \in \mathit{U}, v \in \mathit{I}, \text{ and } \mathbf{1}(u, v) = 1\}$, where $\mathbf{1}(u, v)=1$ if user $u$ interacted with item $v$.
For conversations, $\mathit{R}$ is:
\begin{equation}
    \mathit{R} = \{(u, v) \mid u \in \mathit{U}, \exists t \text{ such that } v \in \mathit{I}^t \text{ and } \mathbf{1}(u, v, t) = 1\}
    \label{eq:interact}
\end{equation}
where $\mathbf{1}(u, v, t)$ indicates if seeker $u$ mentions item $v$ at turn $t$.
However, CRS datasets often lack sufficient user-item interactions for learning robust collaborative signals, as shown in Table~\ref{tab:dataset_stats}~\footnote{Density is calculated as \#Interactions/(\#Users + \#Items)}.
\subsubsection*{\textbf{Linking with Historical Interaction Dataset}}
To address the user-item interactions limitations of conversational datasets, we create a linked historical interaction dataset $\mathit{D}_{CF} = (\mathit{X}, \mathit{V}, \mathit{Z})$ to provide collaborative signals for learning item representations. The dataset consists of three components: (1) $\mathit{X}$ is the set of users in $\mathit{D}_{CF}$, and (2) $\mathit{V}$ is the set of items in $\mathit{D}_{CF}$, with $\mathit{V} \subseteq \mathit{I}$ to ensure alignment with the items in the conversational dataset, and (3) $\mathit{Z}$ is the set of user-item interaction sequences. Each user $x \in \mathit{X}$ has an interaction sequence $Z^x = (i^x_1, i^x_2, \ldots, i^x_k) \in \mathit{Z}$, where $i^x_k \in \mathit{V}$ denotes the $k$-th item interacted with by user $x$. These sequences serve as a foundation for learning collaborative signals to generate rich item representations transferable to CRS.

\subsubsection*{\textbf{Sequential Recommendation for Item Representation Learning}}
The goal of sequential recommendation is to predict the next item in a user interaction sequence based on their historical interactions. Given the interaction sequences $\mathit{Z}$, for a user $x \in \mathit{X}$, let $Z^x_{1:k} = (i^x_1, i^x_2, \ldots, i^x_k)$ be the first $k$ items of the sequence. 
The item embeddings are represented by the matrix $E \in \mathbb{R}^{|V| \times d}$ as:
\begin{equation}
    E^x_{1:k} = (E_{i^x_1}, E_{i^x_2}, \ldots, E_{i^x_k}) \in \mathbb{R}^{k \times d}
    \label{eq:srs_emb}
\end{equation}
This sequence embedding matrix is input into a sequential recommender to predict a next item $i^x_{k+1}$. The training objective is maximizing likelihood of predicting the next item:
\begin{equation}
    \label{eq:obj_sasrec_cf}
    \max_{\Theta} \sum_{x \in \mathit{X}} \sum_{k=1}^{|Z^x|-1} \log p(i^x_{k+1} \mid Z^x_{1:k}; \Theta)
\end{equation}
where $p(i^x_{k+1} \mid Z^x_{1:k}; \Theta)$ represents the probability of the $(k+1)$-th interaction, conditioned on the previous $k$ items, and $\Theta$ is the set of model parameters.
By optimizing Equation~\ref{eq:obj_sasrec_cf}, the model generates item embeddings $E$ that encode collaborative signals. 
Since $\mathit{V} \subseteq \mathit{I}$, embeddings can be directly applied for the conversational dataset to enable effective item ranking and recommendations.
Note that, while our framework focuses on sequential models for generating item embeddings, it is adaptable to non-sequential settings by replacing the backbone model with non-sequential alternatives such as FISM~\cite{kabbur2013fism}. This flexibility allows the framework to support a range of recommendation paradigms.

\subsubsection*{\textbf{Objective}}
The recommender component, following representative works~\cite{li2018towards, chen2019towards, zhou2020improving, he2023large}, optimizes item selection during each conversation turn. During the $k$-th turn, the component exploits the preceding context $(u^{t}, s^{t}, \mathit{I}^{t})_{t=1}^{k-1}$ 
to generate a ranked list $\hat{\mathit{I}}^{k}$, which aligns with ground-truth items in $\mathit{I}^{k}$.

\subsection{Dataset}
We introduce Reddit-ML32M, a novel dataset that bridges Reddit Movie dataset for CRS tasks~\cite{he2023large} with MovieLens 32M, a collaborative filtering dataset~\cite{harper2015movielens}. The key challenge is aligning items across these datasets.
To address this, we use a two-phase linking method. First, IMDb IDs serve as unique identifiers, as they exist in both datasets. However, direct matching may fail when IMDb IDs are modified, reassigned, or deprecated over time, leading to inconsistencies between datasets. To resolve these cases, we use IMDbPY~\footnote{https://github.com/MaximShidlovski23/imdbpy} to retrieve historical IMDb IDs using exact title and release date matching, ensuring correct alignment even when an item's IMDb ID has changed over time or been deprecated.
This linking process enriches the interaction space, increasing total interactions from 51,148 to over 30 million, with a corresponding density improvement from 0.013\% to 0.68\% as shown in Table~\ref{tab:dataset_stats}. We partition the dataset into 80\% training, 10\% validation, and 10\% test sets.

\subsection{Proposed Framework}
We propose \sellname{}, a framework designed to improve CRS, as illustrated in Figure~\ref{fig:teaser} and outlined in Algorithm~\ref{alg:framework}.
Unlike previous models focusing on conversational contexts~\cite{li2018towards, chen2019towards, zhou2020improving, wang2022towards, he2023large, deng2021unified}, we introduce dataset links between CRS and CF modalities to integrate user interaction information from both textual content and item representations generated via implicit feedback. 
Additionally, we incorporate LLMs in a zero-shot setting~\cite{he2023large, friedman2023leveraging} leveraging item representations generated by a pretrained CF recommender as supplementary information.

\subsubsection*{\textbf{Prompting}} 
For each conversation, the context $(u^{t}, s^{t}, \mathit{I}^{t})_{t=1}^{k-1}$ is used to prompt the LLM, with the $k$-th turn being the \textit{recommender's} reply. $s^{t}$ contains raw text for processing by the LLM.
A \texttt{prompt} specifies that the input is intended for conversational recommendation. Pre-trained LLMs are employed in a zero-shot setting to generate item predictions~\cite{radford2019language, he2023large}.
\texttt{prompt} structures the input as:
\begin{equation}
    \label{eq: prompt}
    \texttt{prompt} = \mathit{F}(\texttt{[TASK]}, \texttt{[FORMAT]}, \texttt{[CONTEXT]})
\end{equation}
where \texttt{prompt} is the constructed prompt, \texttt{[TASK]} is a task description template, \texttt{[FORMAT]} is format requirements, and \texttt{[CONTEXT]} is the conversational context. Figure~\ref{fig:teaser} illustrates the transformation.

\subsubsection*{\textbf{LLMs}} 
We employ LLMs in a zero-shot setting to generate recommendations from \texttt{prompt}~\cite{he2023large}. 
For reproducibility, we configure deterministic inference by setting the temperature parameter $\tau=0$, ensuring the softmax selects the highest-probability word at each step~\cite{ackley1985learning}.
The \texttt{response} $=(w_1, w_2, \dots, w_n)$ generated as:
\[
w_i = \arg\max_j \phi_j \quad \text{for } i = 1, 2, \dots, n
\]
where $\phi_j$ is the logit for candidate word $j$ at step $i$. 
This approach leverages the pre-trained knowledge of LLMs without fine-tuning.

\subsubsection*{\textbf{Post-LLM Item Retrieval}} 
Rather than evaluating model weights or output logits from LLMs directly, we implement an exact matching post-processing step that maps a list of recommendations in natural language into a set of in-dataset items $\mathit{I}_{LLM}$.

\subsubsection*{\textbf{Embedding-Based Item Ranking}}
To refine the LLM-generated recommendations $\mathit{I}_{LLM}$, we compute a pairwise similarity matrix between the embeddings of $\mathit{I}_{LLM}$ and all items in the dataset $\mathit{V}$. Using the pretrained embedding matrix $E \in \mathbb{R}^{|V| \times d}$ from SASRec, where $d$ is the embedding dimensionality, we first retrieve the embeddings of the LLM-recommended items, $E_{LLM} \in \mathbb{R}^{m \times d}$.
The pairwise similarity matrix $S \in \mathbb{R}^{m \times |V|}$ is computed as:
\begin{equation}
    \label{eq: sim}
    S_{i,j} = \frac{\mathbf{e}_i \cdot \mathbf{e}_j}{\|\mathbf{e}_i\| \|\mathbf{e}_j\|}, \quad \forall i \in \mathit{I}_{LLM}, \, j \in \mathit{V},
\end{equation}
where $\mathbf{e}_i$ and $\mathbf{e}_j$ are the embeddings of items $i$ and $j$, respectively. The matrix contains similarity scores between each LLM recommended item and all items in $\mathit{V}$. 
We use max pooling across the rows of $S$ to aggregate scores for each item in $\mathit{V}$:
\begin{equation}
    \label{eq: pooling}
    s(j) = \max_{i \in \mathit{I}_{LLM}} S_{i,j}, \quad \forall j \in \mathit{V}.
\end{equation}
This results in a score vector $\mathbf{s} \in \mathbb{R}^{|V|}$, where $s(j)$ represents the maximum similarity of item $j \in \mathit{V}$ with any item in $\mathit{I}_{LLM}$. 
Items in $\mathit{V}$ are ranked based on their scores as $\hat{\mathit{I}}^{k} = \text{argsort}_{j \in \mathit{V}}(-s(j))$.

\begin{algorithm}[t]
\caption{BridgeCRS Algorithm}
\label{alg:framework}
\begin{algorithmic}[1]
\Require Conversation history $C$; pre-trained LLM $\mathcal{M}$; CF embedding matrix $E$; Candidate item set $\mathit{V}$
\Ensure Ranked recommendation list $\hat{\mathit{I}}^k$

\State \textbf{LLM-Based Recommendation}
\State $\texttt{response} \gets \mathcal{M}(\mathit{F}(\texttt{[TASK]}, \texttt{[FORMAT]}, C))$
\State Extract $\mathit{I}_{LLM}$ from $\texttt{response}$

\State \textbf{Post-LLM Item Retrieval}
\State $\mathit{I}_{LLM} \gets \mathit{I}_{LLM} \cap \mathit{V}$ \Comment{Keep only items in candidate set}

\State \textbf{Embedding-Based Item Ranking}
\State Retrieve embeddings $E_{LLM} \in \mathbb{R}^{m \times d}$ for $\mathit{I}_{LLM}$
\State Compute $s(j) = \max_{i \in \mathit{I}_{LLM}} \frac{\mathbf{e}_i \cdot \mathbf{e}_j}{\|\mathbf{e}_i\| \|\mathbf{e}_j\|}, \quad \forall j \in \mathit{V}$
\State \Return $\hat{\mathit{I}}^k = \text{argsort}_{j \in \mathit{V}}(-s(j))$
\end{algorithmic}
\end{algorithm}

\section{Experiments}

\begin{table*}[htbp]

\caption{Comparison of performance against (1) CF-based recommenders, (2) CRS models, and (3) LLM-based CRS without CF signals for k = \{1,5,10\}. Results include percentage improvements over the best underlined values, with best values in bold.}
% \vspace{-10pt}
\label{tab:results}
\centering
{
\small
\begin{tabular}{@{}cccccccccc@{}}
\toprule

\textbf{Models} & H@1 & N@1 & H@5 & N@5 & H@10 & N@10 \\ 
\midrule

\textbf{PopRec} 
    & .0022 ± .0004 & .0022 ± .0004 & .0087 ± .0011 & .0046 ± .0004 & .0131 ± .0011 & .0064 ± .0005  \\ 

\textbf{item2vec} 
    & .0078 ± .0009 & .0078 ± .0009 & .0274 ± .0016 & .0131 ± .0006 & .0436 ± .0020 & .0172 ± .0017  \\ 

\textbf{FISM} 
    & .0046 ± .0007 & .0046 ± .0007 & .0229 ± .0015  & .0085 ± .0016 & .0441 ± .0021 & .0132 ± .0029  \\  

\textbf{SASRec} 
    & .0059 ± .0014 & .0059 ± .0014 & .0270 ± .0010 & .0130 ± .0010 & .0390 ± .0010 & .0180 ± .0010 \\ 

\midrule

\textbf{ReDIAL}

    % &  &  &  &  &  &  &  &  &  \\ 
    & .0050 ± .0010 & - & .0290 ± .0010 & .0190 ± .0010 & .0440 ± .0010 & .0240 ± .0010 \\ 

\textbf{UniCRS} 
    & .0020 ± .0006 & .0020 ± .0006 & .0260 ± .0010 & .0170 ± .0010 & .0430 ± .0010 &  .0220 ± .0010\\ 

\midrule

\textbf{Llama2}
    & - & - & .0420 ± .0010 & .0270 ± .0010 & .064 ± .0010 & .0340 ± .0010 \\ 

\textbf{\gpt{}} 
    & \underline{.0152 ± .0012}  & \underline{.0152 ± .0012} & \underline{.0686 ± .0024} & \underline{.0289 ± .0012} & \underline{.1218 ± .0031} & \underline{.0419 ± .0013} \\

\midrule

\textbf{\sellname{} (Ours)}
    & \textbf{.0156 ± .0012} & \textbf{.0156 ± .0012} & \textbf{.0770 ± .0025} & \textbf{.0318 ± .0012} & \textbf{.1330 ± .0032} & \textbf{.0449 ± .0013} \\

\midrule

\textbf{Improvement (\%)}
    & +2.96\% & +2.96\% & +12.32\% &  +9.91\% & +9.14\% & +7.22\% \\

\bottomrule
\end{tabular}
}
% \vspace{-10pt}
\end{table*}

\subsubsection*{\textbf{Experimental Setup}}
Our framework integrates embeddings generated by SASRec \cite{kang2018self} for collaborative filtering with conversational recommendations generated by \gpt{} to create a unified CRS that leverages the strengths of both approaches. 
To assess performance, we evaluate our framework against three groups of representative baseline models.
First, traditional item-based recommendation models include popularity (PopRec), FISM~\cite{kabbur2013fism}, item2vec~\cite{barkan2016item2vec}, and SASRec~\cite{kang2018self}, trained on interactions from the Reddit Movie dataset, as shown in Table~\ref{tab:dataset_stats}. These models serve as representative baselines for assessing recommendation performance.
Second, traditional CRS models include ReDIAL\footnotemark[4]\cite{li2018towards} and UniCRS~\cite{wang2022towards}. UniCRS leverages pre-trained language models with prompt tuning to generate recommendations. We do not other works~\cite{li2022user, ma-etal-2021-cr}, as UniCRS is representative with similar results.
Third, zero-shot LLM-based CRS models, including LLaMA2-7B\footnotemark[4]\cite{touvron2023llama} and \gpt{}, which generate conversational recommendations without fine-tuning~\cite{he2023large}. \gpt{} stands out as a state-of-the-art model for CRS tasks, as noted in~\cite{he2024reindex} under the zero-shot setting defined in~\cite{he2023large}. Our experiments confirm that \gpt{} achieves the best results across key metrics, validating its position as a leading baseline for CRS.

\footnotetext[4]{Results as reported in~\cite{he2023large, he2024reindex}}

\subsubsection*{\textbf{Evaluation Protocol}}
Following prior studies~\cite{li2018towards, chen2019towards, zhou2020improving, he2023large, wang2022towards, he2024reindex}, we generate predictions for each conversation and select the top $k$ recommendations. We evaluate against ground truth items $\hat{\mathit{I}}^{k}$. We report Hit Rate (H@$k$) and NDCG@$k$ (N@$k$) for $K=\{1,5,10\}$.

\subsection{Results and Analysis}
We assess performance of our framework against eight baselines as shown in Table~\ref{tab:results}. Our method consistently outperforms all baselines, demonstrating the effectiveness of integrating CF signals into conversational recommendations. Specifically, our framework achieves a 12.32\% improvement in H@5 over the state-of-the-art baseline~\cite{he2024reindex}, \gpt{}. Similarly, N@5 improves by 9.91\%, further confirming the ranking quality of our method. Note that H@1 and N@1 are identical because NDCG@1 reduces to Hit@1 due to the absence of rank discounting.

CF-based models perform well in leveraging historical user-item interactions but fall short in conversational settings due to their inability to incorporate contextual information. In contrast, CRS models such as ReDIAL and UniCRS incorporate conversational data but are limited in their ability to capture behavioral patterns. 
LLM-based CRS models leverage conversational context effectively but struggle with less frequent items due to the absence of item representations.
Our framework addresses these limitations by using CF embeddings with LLM-generated recommendations, resulting in superior accuracy across all scenarios. Unlike SASRec, which relies on historical interactions, or \gpt{}, which lacks item-level embeddings, our framework bridges the gap between CF and conversational understanding.

\subsubsection*{\textbf{Generalization to Other Domains}}
The proposed dataset linking method is broadly applicable across domains where structured identifiers facilitate dataset alignment, such as ASINs in e-commerce and ISBNs in books~\cite{ni19justifying, wan18item}.
This approach extends to industry settings where proprietary user interactions contain unique product identifiers, allowing collaborative item representations to be generated from implicit feedback sources such as purchase history and browsing patterns~\cite{yoon2024forecasting}. By leveraging structured identifiers and user interaction data, this framework can integrate collaborative signals into CRS tasks, which bridging structured and conversational modalities across diverse domains.

\section{Related Work}
Exploiting collaborative signals can be achieved using matrix factorization~\cite{koren2009matrix} or a Graph Convolutional Network (GCN)~\cite{he2020lightgcn}. SASRec~\cite{kang2018self} uses a self-attention network to exploit sequential patterns within collaborative signals. Using conversational context typically involves leveraging the semantics of conversational text and recommending items to maximize alignment between conversation and item attributes. ReDial~\cite{li2018towards} predicts user sentiment in the conversational context and feeds it to an autoencoder-based recommendation. The autoencoder-based recommender is similar to AutoRec~\cite{sedhain2015autorec} which is trained to produce unobserved user ratings from the observed user ratings. EAR~\cite{lei2020estimation} extracts item attributes and attribute-specific user preferences from the dialogue context and feeds them to an attribute-aware Bayesian Personalized Ranking (BPR) model for recommendation. KBRD~\cite{chen2019towards} extracts item and non-item entities from the conversation and feeds them through a Relational Graph Convolutional Network (RGCN)~\cite{schlichtkrull2018modeling} to generate a knowledge-enhanced user preference representation. The RGCN is trained with an item-oriented knowledge graph from DBpedia~\cite{lehmann2015dbpedia}. KGSF~\cite{zhou2020improving} exploits a word-oriented knowledge graph via a GCN in addition to item-oriented knowledge to recommend items based on keywords in the conversation. The GCN associated with the word-oriented knowledge graph is trained on ConceptNet~\cite{speer20165}. UniCRS~\cite{wang2022towards} jointly optimizes entity embeddings from an RGCN and contextual word embeddings from pretrained language models. 

Recent works have explored the direct use of conversational context by prompting LLMs~\cite{friedman2023leveraging,hou2024llmrank}. He et al.~\cite{he2023large} feed dialogue context directly to LLMs for generating recommendations and do not use user-item interaction data. Liu et al.~\cite{liu2023chatgpt} demonstrate the use of LLMs for several recommendation tasks such as direct recommendation, rating prediction, sequential recommendation, explanation generation, and review summarization by constructing task-specific prompts. However, this approach performs poorly on direct and sequential recommendation~\cite{zhang2023recommendationinstructionfollowinglarge}. Gao et al.~\cite{gao2023chat} propose adding user-item interaction history to the LLM prompts in addition to the dialogue context. However, LLMs cannot exploit complex user-item interactions from their textual representations. ILM~\cite{yang2024item} uses item-text contrastive learning for learning text-aligned item representations from collaborative signals. They train an item encoder to produce text-aligned item representations so that a frozen LLM may exploit item representations with existing parametric knowledge. Zhang et al.~\cite{zhang2023collm} propose encoding user-item interaction using traditional collaborative models and tuning a low-rank adapter associated with LLM so that collaborative information can be exploited by LLMs. TALLRec~\cite{Bao_2023} proposes instruction tuning of LLMs to improve their ability to exploit collaborative signals.
\section{Conclusion and Future Directions}
We present the Reddit-ML32M dataset and a framework that integrates CF signals with conversational context to improve CRS. Our approach achieves consistent performance improvements, with a 12.32\% increase in recall and a 9.91\% improvement in NDCG compared to existing methods.
The dataset serves as a resource for research in CRS, while our methodology shows promising future directions for constructing diverse datasets across different domains. This opens opportunities for broader CRS applications and tackling interaction sparsity in future work.

\bibliographystyle{ACM-Reference-Format}
\bibliography{references}


\begin{thebibliography}{42}

%%% the Web address.

\ifx \showCODEN    \undefined \def \showCODEN     #1{\unskip}     \fi
\ifx \showDOI      \undefined \def \showDOI       #1{#1}\fi
\ifx \showISBNx    \undefined \def \showISBNx     #1{\unskip}     \fi
\ifx \showISBNxiii \undefined \def \showISBNxiii  #1{\unskip}     \fi
\ifx \showISSN     \undefined \def \showISSN      #1{\unskip}     \fi
\ifx \showLCCN     \undefined \def \showLCCN      #1{\unskip}     \fi
\ifx \shownote     \undefined \def \shownote      #1{#1}          \fi
\ifx \showarticletitle \undefined \def \showarticletitle #1{#1}   \fi
\ifx \showURL      \undefined \def \showURL       {\relax}        \fi
% invisible to TeX
\providecommand\bibfield[2]{#2}
\providecommand\bibinfo[2]{#2}
\providecommand\natexlab[1]{#1}
\providecommand\showeprint[2][]{arXiv:#2}

\bibitem[Ackley et~al\mbox{.}(1985)]%
        {ackley1985learning}
\bibfield{author}{\bibinfo{person}{David~H Ackley}, \bibinfo{person}{Geoffrey~E Hinton}, {and} \bibinfo{person}{Terrence~J Sejnowski}.} \bibinfo{year}{1985}\natexlab{}.
\newblock \showarticletitle{A learning algorithm for Boltzmann machines}.
\newblock \bibinfo{journal}{\emph{Cognitive science}} \bibinfo{volume}{9}, \bibinfo{number}{1} (\bibinfo{year}{1985}), \bibinfo{pages}{147--169}.
\newblock


\bibitem[Bao et~al\mbox{.}(2023)]%
        {Bao_2023}
\bibfield{author}{\bibinfo{person}{Keqin Bao}, \bibinfo{person}{Jizhi Zhang}, \bibinfo{person}{Yang Zhang}, \bibinfo{person}{Wenjie Wang}, \bibinfo{person}{Fuli Feng}, {and} \bibinfo{person}{Xiangnan He}.} \bibinfo{year}{2023}\natexlab{}.
\newblock \showarticletitle{TALLRec: An Effective and Efficient Tuning Framework to Align Large Language Model with Recommendation}. In \bibinfo{booktitle}{\emph{RecSys}}.
\newblock


\bibitem[Barkan and Koenigstein(2016)]%
        {barkan2016item2vec}
\bibfield{author}{\bibinfo{person}{Oren Barkan} {and} \bibinfo{person}{Noam Koenigstein}.} \bibinfo{year}{2016}\natexlab{}.
\newblock \showarticletitle{{ITEM2VEC:} Neural item embedding for collaborative filtering}. In \bibinfo{booktitle}{\emph{26th {IEEE} International Workshop on Machine Learning for Signal Processing, {MLSP}}}. \bibinfo{publisher}{{IEEE}}.
\newblock


\bibitem[Brown et~al\mbox{.}(2020)]%
        {brown2020language}
\bibfield{author}{\bibinfo{person}{Tom Brown}, \bibinfo{person}{Benjamin Mann}, \bibinfo{person}{Nick Ryder}, \bibinfo{person}{Melanie Subbiah}, \bibinfo{person}{Jared~D Kaplan}, \bibinfo{person}{Prafulla Dhariwal}, \bibinfo{person}{Arvind Neelakantan}, \bibinfo{person}{Pranav Shyam}, \bibinfo{person}{Girish Sastry}, \bibinfo{person}{Amanda Askell}, {et~al\mbox{.}}} \bibinfo{year}{2020}\natexlab{}.
\newblock \showarticletitle{Language models are few-shot learners}.
\newblock \bibinfo{journal}{\emph{Advances in neural information processing systems}}  \bibinfo{volume}{33} (\bibinfo{year}{2020}), \bibinfo{pages}{1877--1901}.
\newblock


\bibitem[Chen et~al\mbox{.}(2019)]%
        {chen2019towards}
\bibfield{author}{\bibinfo{person}{Qibin Chen}, \bibinfo{person}{Junyang Lin}, \bibinfo{person}{Yichang Zhang}, \bibinfo{person}{Ming Ding}, \bibinfo{person}{Yukuo Cen}, \bibinfo{person}{Hongxia Yang}, {and} \bibinfo{person}{Jie Tang}.} \bibinfo{year}{2019}\natexlab{}.
\newblock \showarticletitle{Towards Knowledge-Based Recommender Dialog System}. In \bibinfo{booktitle}{\emph{EMNLP-IJCNLP}}.
\newblock


\bibitem[Chowdhery et~al\mbox{.}(2023)]%
        {chowdhery2023palm}
\bibfield{author}{\bibinfo{person}{Aakanksha Chowdhery}, \bibinfo{person}{Sharan Narang}, \bibinfo{person}{Jacob Devlin}, \bibinfo{person}{Maarten Bosma}, \bibinfo{person}{Gaurav Mishra}, \bibinfo{person}{Adam Roberts}, \bibinfo{person}{Paul Barham}, \bibinfo{person}{Hyung~Won Chung}, \bibinfo{person}{Charles Sutton}, \bibinfo{person}{Sebastian Gehrmann}, {et~al\mbox{.}}} \bibinfo{year}{2023}\natexlab{}.
\newblock \showarticletitle{Palm: Scaling language modeling with pathways}.
\newblock \bibinfo{journal}{\emph{Journal of Machine Learning Research}} \bibinfo{volume}{24}, \bibinfo{number}{240} (\bibinfo{year}{2023}), \bibinfo{pages}{1--113}.
\newblock


\bibitem[Deng et~al\mbox{.}(2021)]%
        {deng2021unified}
\bibfield{author}{\bibinfo{person}{Yang Deng}, \bibinfo{person}{Yaliang Li}, \bibinfo{person}{Fei Sun}, \bibinfo{person}{Bolin Ding}, {and} \bibinfo{person}{Wai Lam}.} \bibinfo{year}{2021}\natexlab{}.
\newblock \showarticletitle{Unified Conversational Recommendation Policy Learning via Graph-based Reinforcement Learning}. In \bibinfo{booktitle}{\emph{Proceedings of the 44th International ACM SIGIR Conference on Research and Development in Information Retrieval}} \emph{(\bibinfo{series}{SIGIR '21})}. \bibinfo{publisher}{Association for Computing Machinery}, \bibinfo{pages}{1431–1441}.
\newblock
\showISBNx{9781450380379}


\bibitem[Friedman et~al\mbox{.}(2023)]%
        {friedman2023leveraging}
\bibfield{author}{\bibinfo{person}{Luke Friedman}, \bibinfo{person}{Sameer Ahuja}, \bibinfo{person}{David Allen}, \bibinfo{person}{Terry Tan}, \bibinfo{person}{Hakim Sidahmed}, \bibinfo{person}{Changbo Long}, \bibinfo{person}{Jun Xie}, \bibinfo{person}{Gabriel Schubiner}, \bibinfo{person}{Ajay Patel}, \bibinfo{person}{Harsh Lara}, {et~al\mbox{.}}} \bibinfo{year}{2023}\natexlab{}.
\newblock \showarticletitle{Leveraging large language models in conversational recommender systems}.
\newblock \bibinfo{journal}{\emph{arXiv preprint arXiv:2305.07961}} (\bibinfo{year}{2023}).
\newblock


\bibitem[Gao et~al\mbox{.}(2023)]%
        {gao2023chat}
\bibfield{author}{\bibinfo{person}{Yunfan Gao}, \bibinfo{person}{Tao Sheng}, \bibinfo{person}{Youlin Xiang}, \bibinfo{person}{Yun Xiong}, \bibinfo{person}{Haofen Wang}, {and} \bibinfo{person}{Jiawei Zhang}.} \bibinfo{year}{2023}\natexlab{}.
\newblock \bibinfo{title}{Chat-REC: Towards Interactive and Explainable LLMs-Augmented Recommender System}.
\newblock
\newblock
\showeprint[arxiv]{2303.14524}~[cs.IR]
\urldef\tempurl%
\url{https://arxiv.org/abs/2303.14524}
\showURL{%
\tempurl}


\bibitem[Harper and Konstan(2015)]%
        {harper2015movielens}
\bibfield{author}{\bibinfo{person}{F~Maxwell Harper} {and} \bibinfo{person}{Joseph~A Konstan}.} \bibinfo{year}{2015}\natexlab{}.
\newblock \showarticletitle{The movielens datasets: History and context}.
\newblock \bibinfo{journal}{\emph{Acm transactions on interactive intelligent systems (tiis)}} \bibinfo{volume}{5}, \bibinfo{number}{4} (\bibinfo{year}{2015}).
\newblock


\bibitem[He et~al\mbox{.}(2020)]%
        {he2020lightgcn}
\bibfield{author}{\bibinfo{person}{Xiangnan He}, \bibinfo{person}{Kuan Deng}, \bibinfo{person}{Xiang Wang}, \bibinfo{person}{Yan Li}, \bibinfo{person}{YongDong Zhang}, {and} \bibinfo{person}{Meng Wang}.} \bibinfo{year}{2020}\natexlab{}.
\newblock \showarticletitle{LightGCN: Simplifying and Powering Graph Convolution Network for Recommendation}. In \bibinfo{booktitle}{\emph{Proceedings of the 43rd International ACM SIGIR Conference on Research and Development in Information Retrieval}} \emph{(\bibinfo{series}{SIGIR '20})}.
\newblock


\bibitem[He et~al\mbox{.}(2023)]%
        {he2023large}
\bibfield{author}{\bibinfo{person}{Zhankui He}, \bibinfo{person}{Zhouhang Xie}, \bibinfo{person}{Rahul Jha}, \bibinfo{person}{Harald Steck}, \bibinfo{person}{Dawen Liang}, \bibinfo{person}{Yesu Feng}, \bibinfo{person}{Bodhisattwa~Prasad Majumder}, \bibinfo{person}{Nathan Kallus}, {and} \bibinfo{person}{Julian McAuley}.} \bibinfo{year}{2023}\natexlab{}.
\newblock \showarticletitle{Large language models as zero-shot conversational recommenders}. In \bibinfo{booktitle}{\emph{CIKM}}.
\newblock


\bibitem[He et~al\mbox{.}(2025)]%
        {he2024reindex}
\bibfield{author}{\bibinfo{person}{Zhankui He}, \bibinfo{person}{Zhouhang Xie}, \bibinfo{person}{Harald Steck}, \bibinfo{person}{Dawen Liang}, \bibinfo{person}{Rahul Jha}, \bibinfo{person}{Nathan Kallus}, {and} \bibinfo{person}{Julian McAuley}.} \bibinfo{year}{2025}\natexlab{}.
\newblock \showarticletitle{Reindex-then-adapt: Improving large language models for conversational recommendation}. In \bibinfo{booktitle}{\emph{WSDM}}.
\newblock


\bibitem[Hou et~al\mbox{.}(2024)]%
        {hou2024llmrank}
\bibfield{author}{\bibinfo{person}{Yupeng Hou}, \bibinfo{person}{Junjie Zhang}, \bibinfo{person}{Zihan Lin}, \bibinfo{person}{Hongyu Lu}, \bibinfo{person}{Ruobing Xie}, \bibinfo{person}{Julian McAuley}, {and} \bibinfo{person}{Wayne~Xin Zhao}.} \bibinfo{year}{2024}\natexlab{}.
\newblock \showarticletitle{Large Language Models are Zero-Shot Rankers for Recommender Systems}. In \bibinfo{booktitle}{\emph{{ECIR}}}.
\newblock


\bibitem[Kabbur et~al\mbox{.}(2013)]%
        {kabbur2013fism}
\bibfield{author}{\bibinfo{person}{Santosh Kabbur}, \bibinfo{person}{Xia Ning}, {and} \bibinfo{person}{George Karypis}.} \bibinfo{year}{2013}\natexlab{}.
\newblock \showarticletitle{Fism: factored item similarity models for top-n recommender systems}. In \bibinfo{booktitle}{\emph{KDD}}.
\newblock


\bibitem[Kang and McAuley(2018)]%
        {kang2018self}
\bibfield{author}{\bibinfo{person}{Wang-Cheng Kang} {and} \bibinfo{person}{Julian McAuley}.} \bibinfo{year}{2018}\natexlab{}.
\newblock \showarticletitle{Self-attentive sequential recommendation}. In \bibinfo{booktitle}{\emph{2018 IEEE international conference on data mining (ICDM)}}. IEEE.
\newblock


\bibitem[Kim et~al\mbox{.}(2024)]%
        {kim2024large}
\bibfield{author}{\bibinfo{person}{Sein Kim}, \bibinfo{person}{Hongseok Kang}, \bibinfo{person}{Seungyoon Choi}, \bibinfo{person}{Donghyun Kim}, \bibinfo{person}{Minchul Yang}, {and} \bibinfo{person}{Chanyoung Park}.} \bibinfo{year}{2024}\natexlab{}.
\newblock \showarticletitle{Large language models meet collaborative filtering: An efficient all-round llm-based recommender system}. In \bibinfo{booktitle}{\emph{KDD}}.
\newblock


\bibitem[Koren et~al\mbox{.}(2009)]%
        {koren2009matrix}
\bibfield{author}{\bibinfo{person}{Yehuda Koren}, \bibinfo{person}{Robert Bell}, {and} \bibinfo{person}{Chris Volinsky}.} \bibinfo{year}{2009}\natexlab{}.
\newblock \showarticletitle{Matrix factorization techniques for recommender systems}.
\newblock \bibinfo{journal}{\emph{Computer}} \bibinfo{volume}{42}, \bibinfo{number}{8} (\bibinfo{year}{2009}), \bibinfo{pages}{30--37}.
\newblock


\bibitem[Lehmann et~al\mbox{.}(2015)]%
        {lehmann2015dbpedia}
\bibfield{author}{\bibinfo{person}{Jens Lehmann}, \bibinfo{person}{Robert Isele}, \bibinfo{person}{Max Jakob}, \bibinfo{person}{Anja Jentzsch}, \bibinfo{person}{Dimitris Kontokostas}, \bibinfo{person}{Pablo~N. Mendes}, \bibinfo{person}{Sebastian Hellmann}, \bibinfo{person}{Mohamed Morsey}, \bibinfo{person}{Patrick van Kleef}, \bibinfo{person}{S{\"{o}}ren Auer}, {and} \bibinfo{person}{Christian Bizer}.} \bibinfo{year}{2015}\natexlab{}.
\newblock \showarticletitle{DBpedia - {A} large-scale, multilingual knowledge base extracted from Wikipedia}.
\newblock \bibinfo{journal}{\emph{Semantic Web}} (\bibinfo{year}{2015}).
\newblock


\bibitem[Lei et~al\mbox{.}(2020)]%
        {lei2020estimation}
\bibfield{author}{\bibinfo{person}{Wenqiang Lei}, \bibinfo{person}{Xiangnan He}, \bibinfo{person}{Yisong Miao}, \bibinfo{person}{Qingyun Wu}, \bibinfo{person}{Richang Hong}, \bibinfo{person}{Min-Yen Kan}, {and} \bibinfo{person}{Tat-Seng Chua}.} \bibinfo{year}{2020}\natexlab{}.
\newblock \showarticletitle{Estimation-Action-Reflection: Towards Deep Interaction Between Conversational and Recommender Systems}. In \bibinfo{booktitle}{\emph{WSDM '20}}.
\newblock


\bibitem[Li et~al\mbox{.}(2018)]%
        {li2018towards}
\bibfield{author}{\bibinfo{person}{Raymond Li}, \bibinfo{person}{Samira Ebrahimi~Kahou}, \bibinfo{person}{Hannes Schulz}, \bibinfo{person}{Vincent Michalski}, \bibinfo{person}{Laurent Charlin}, {and} \bibinfo{person}{Chris Pal}.} \bibinfo{year}{2018}\natexlab{}.
\newblock \showarticletitle{Towards deep conversational recommendations}.
\newblock \bibinfo{journal}{\emph{NeurIPS}} (\bibinfo{year}{2018}).
\newblock


\bibitem[Li et~al\mbox{.}(2022)]%
        {li2022user}
\bibfield{author}{\bibinfo{person}{Shuokai Li}, \bibinfo{person}{Ruobing Xie}, \bibinfo{person}{Yongchun Zhu}, \bibinfo{person}{Xiang Ao}, \bibinfo{person}{Fuzhen Zhuang}, {and} \bibinfo{person}{Qing He}.} \bibinfo{year}{2022}\natexlab{}.
\newblock \showarticletitle{User-centric conversational recommendation with multi-aspect user modeling}. In \bibinfo{booktitle}{\emph{SIGIR}}.
\newblock


\bibitem[Liu et~al\mbox{.}(2023)]%
        {liu2023chatgpt}
\bibfield{author}{\bibinfo{person}{Junling Liu}, \bibinfo{person}{Chao Liu}, \bibinfo{person}{Peilin Zhou}, \bibinfo{person}{Renjie Lv}, \bibinfo{person}{Kang Zhou}, {and} \bibinfo{person}{Yan Zhang}.} \bibinfo{year}{2023}\natexlab{}.
\newblock \bibinfo{title}{Is ChatGPT a Good Recommender? A Preliminary Study}.
\newblock
\newblock
\showeprint[arxiv]{2304.10149}~[cs.IR]
\urldef\tempurl%
\url{https://arxiv.org/abs/2304.10149}
\showURL{%
\tempurl}


\bibitem[Ma et~al\mbox{.}(2021)]%
        {ma-etal-2021-cr}
\bibfield{author}{\bibinfo{person}{Wenchang Ma}, \bibinfo{person}{Ryuichi Takanobu}, {and} \bibinfo{person}{Minlie Huang}.} \bibinfo{year}{2021}\natexlab{}.
\newblock \showarticletitle{{CR}-Walker: Tree-Structured Graph Reasoning and Dialog Acts for Conversational Recommendation}. In \bibinfo{booktitle}{\emph{EMNLP}}.
\newblock


\bibitem[Ni et~al\mbox{.}(2019)]%
        {ni19justifying}
\bibfield{author}{\bibinfo{person}{Jianmo Ni}, \bibinfo{person}{Jiacheng Li}, {and} \bibinfo{person}{Julian McAuley}.} \bibinfo{year}{2019}\natexlab{}.
\newblock \showarticletitle{Justifying recommendations using distantly-labeled reviews and fined-grained aspects}. In \bibinfo{booktitle}{\emph{EMNLP}}.
\newblock


\bibitem[OpenAI(2024)]%
        {gpt-models}
\bibfield{author}{\bibinfo{person}{OpenAI}.} \bibinfo{year}{2024}\natexlab{}.
\newblock \bibinfo{title}{Models - OpenAI API}.
\newblock \bibinfo{howpublished}{\url{https://platform.openai.com/docs/models}}.
\newblock
\newblock
\shownote{Accessed: 2024-06-13}.


\bibitem[Ouyang et~al\mbox{.}(2022)]%
        {ouyang2022training}
\bibfield{author}{\bibinfo{person}{Long Ouyang}, \bibinfo{person}{Jeffrey Wu}, \bibinfo{person}{Xu Jiang}, \bibinfo{person}{Diogo Almeida}, \bibinfo{person}{Carroll Wainwright}, \bibinfo{person}{Pamela Mishkin}, \bibinfo{person}{Chong Zhang}, \bibinfo{person}{Sandhini Agarwal}, \bibinfo{person}{Katarina Slama}, \bibinfo{person}{Alex Ray}, {et~al\mbox{.}}} \bibinfo{year}{2022}\natexlab{}.
\newblock \showarticletitle{Training language models to follow instructions with human feedback}.
\newblock \bibinfo{journal}{\emph{Advances in neural information processing systems}}  \bibinfo{volume}{35} (\bibinfo{year}{2022}), \bibinfo{pages}{27730--27744}.
\newblock


\bibitem[Radford et~al\mbox{.}(2019)]%
        {radford2019language}
\bibfield{author}{\bibinfo{person}{Alec Radford}, \bibinfo{person}{Jeffrey Wu}, \bibinfo{person}{Rewon Child}, \bibinfo{person}{David Luan}, \bibinfo{person}{Dario Amodei}, \bibinfo{person}{Ilya Sutskever}, {et~al\mbox{.}}} \bibinfo{year}{2019}\natexlab{}.
\newblock \showarticletitle{Language models are unsupervised multitask learners}.
\newblock \bibinfo{journal}{\emph{OpenAI blog}} \bibinfo{volume}{1}, \bibinfo{number}{8} (\bibinfo{year}{2019}), \bibinfo{pages}{9}.
\newblock


\bibitem[Salem et~al\mbox{.}(2014)]%
        {salem2014history}
\bibfield{author}{\bibinfo{person}{Yasser Salem}, \bibinfo{person}{Jun Hong}, {and} \bibinfo{person}{Weiru Liu}.} \bibinfo{year}{2014}\natexlab{}.
\newblock \showarticletitle{History-guided conversational recommendation}. In \bibinfo{booktitle}{\emph{WWW 14}}.
\newblock


\bibitem[Schlichtkrull et~al\mbox{.}(2018)]%
        {schlichtkrull2018modeling}
\bibfield{author}{\bibinfo{person}{Michael Schlichtkrull}, \bibinfo{person}{Thomas~N. Kipf}, \bibinfo{person}{Peter Bloem}, \bibinfo{person}{Rianne van\&nbsp;den Berg}, \bibinfo{person}{Ivan Titov}, {and} \bibinfo{person}{Max Welling}.} \bibinfo{year}{2018}\natexlab{}.
\newblock \showarticletitle{Modeling Relational Data with Graph Convolutional Networks}. In \bibinfo{booktitle}{\emph{The Semantic Web: 15th International Conference, ESWC}}.
\newblock


\bibitem[Sedhain et~al\mbox{.}(2015)]%
        {sedhain2015autorec}
\bibfield{author}{\bibinfo{person}{Suvash Sedhain}, \bibinfo{person}{Aditya~Krishna Menon}, \bibinfo{person}{Scott Sanner}, {and} \bibinfo{person}{Lexing Xie}.} \bibinfo{year}{2015}\natexlab{}.
\newblock \showarticletitle{AutoRec: Autoencoders Meet Collaborative Filtering}. In \bibinfo{booktitle}{\emph{Proceedings of the 24th International Conference on World Wide Web}} \emph{(\bibinfo{series}{WWW '15 Companion})}.
\newblock


\bibitem[Speer et~al\mbox{.}(2017)]%
        {speer20165}
\bibfield{author}{\bibinfo{person}{Robyn Speer}, \bibinfo{person}{Joshua Chin}, {and} \bibinfo{person}{Catherine Havasi}.} \bibinfo{year}{2017}\natexlab{}.
\newblock \showarticletitle{ConceptNet 5.5: an open multilingual graph of general knowledge}. In \bibinfo{booktitle}{\emph{AAAI}}.
\newblock


\bibitem[Sun and Zhang(2018)]%
        {sun2018conversational}
\bibfield{author}{\bibinfo{person}{Yueming Sun} {and} \bibinfo{person}{Yi Zhang}.} \bibinfo{year}{2018}\natexlab{}.
\newblock \showarticletitle{Conversational recommender system}. In \bibinfo{booktitle}{\emph{The 41st international acm sigir conference on research \& development in information retrieval}}. \bibinfo{pages}{235--244}.
\newblock


\bibitem[Touvron et~al\mbox{.}(2023)]%
        {touvron2023llama}
\bibfield{author}{\bibinfo{person}{Hugo Touvron}, \bibinfo{person}{Louis Martin}, \bibinfo{person}{Kevin Stone}, \bibinfo{person}{Peter Albert}, \bibinfo{person}{Amjad Almahairi}, \bibinfo{person}{Yasmine Babaei}, \bibinfo{person}{Nikolay Bashlykov}, \bibinfo{person}{Soumya Batra}, \bibinfo{person}{Prajjwal Bhargava}, \bibinfo{person}{Shruti Bhosale}, {et~al\mbox{.}}} \bibinfo{year}{2023}\natexlab{}.
\newblock \showarticletitle{Llama 2: Open foundation and fine-tuned chat models}.
\newblock \bibinfo{journal}{\emph{arXiv preprint arXiv:2307.09288}} (\bibinfo{year}{2023}).
\newblock


\bibitem[Wan and McAuley(2018)]%
        {wan18item}
\bibfield{author}{\bibinfo{person}{Mengting Wan} {and} \bibinfo{person}{Julian McAuley}.} \bibinfo{year}{2018}\natexlab{}.
\newblock \showarticletitle{Item recommendation on monotonic behavior chains}. In \bibinfo{booktitle}{\emph{RecSys}}.
\newblock


\bibitem[Wang et~al\mbox{.}(2023)]%
        {wang-etal-2023-rethinking-evaluation}
\bibfield{author}{\bibinfo{person}{Xiaolei Wang}, \bibinfo{person}{Xinyu Tang}, \bibinfo{person}{Xin Zhao}, \bibinfo{person}{Jingyuan Wang}, {and} \bibinfo{person}{Ji-Rong Wen}.} \bibinfo{year}{2023}\natexlab{}.
\newblock \showarticletitle{Rethinking the Evaluation for Conversational Recommendation in the Era of Large Language Models}. In \bibinfo{booktitle}{\emph{Proceedings of the 2023 Conference on Empirical Methods in Natural Language Processing}}.
\newblock


\bibitem[Wang et~al\mbox{.}(2022)]%
        {wang2022towards}
\bibfield{author}{\bibinfo{person}{Xiaolei Wang}, \bibinfo{person}{Kun Zhou}, \bibinfo{person}{Ji-Rong Wen}, {and} \bibinfo{person}{Wayne~Xin Zhao}.} \bibinfo{year}{2022}\natexlab{}.
\newblock \showarticletitle{Towards unified conversational recommender systems via knowledge-enhanced prompt learning}. In \bibinfo{booktitle}{\emph{KDD}}.
\newblock


\bibitem[Yang et~al\mbox{.}(2024)]%
        {yang2024item}
\bibfield{author}{\bibinfo{person}{Li Yang}, \bibinfo{person}{Anushya Subbiah}, \bibinfo{person}{Hardik Patel}, \bibinfo{person}{Judith~Yue Li}, \bibinfo{person}{Yanwei Song}, \bibinfo{person}{Reza Mirghaderi}, {and} \bibinfo{person}{Vikram Aggarwal}.} \bibinfo{year}{2024}\natexlab{}.
\newblock \showarticletitle{Item-Language Model for Conversational Recommendation}.
\newblock \bibinfo{journal}{\emph{arXiv preprint arXiv:2406.02844}} (\bibinfo{year}{2024}).
\newblock


\bibitem[Yoon et~al\mbox{.}(2024)]%
        {yoon2024forecasting}
\bibfield{author}{\bibinfo{person}{Se-eun Yoon}, \bibinfo{person}{Ahmad Bin~Rabiah}, \bibinfo{person}{Zaid Alibadi}, \bibinfo{person}{Surya Kallumadi}, {and} \bibinfo{person}{Julian McAuley}.} \bibinfo{year}{2024}\natexlab{}.
\newblock \showarticletitle{Forecasting Live Chat Intent from Browsing History}. In \bibinfo{booktitle}{\emph{CIKM}}.
\newblock


\bibitem[Zhang et~al\mbox{.}(2023b)]%
        {zhang2023recommendationinstructionfollowinglarge}
\bibfield{author}{\bibinfo{person}{Junjie Zhang}, \bibinfo{person}{Ruobing Xie}, \bibinfo{person}{Yupeng Hou}, \bibinfo{person}{Wayne~Xin Zhao}, \bibinfo{person}{Leyu Lin}, {and} \bibinfo{person}{Ji-Rong Wen}.} \bibinfo{year}{2023}\natexlab{b}.
\newblock \bibinfo{title}{Recommendation as Instruction Following: A Large Language Model Empowered Recommendation Approach}.
\newblock
\newblock
\showeprint[arxiv]{2305.07001}~[cs.IR]
\urldef\tempurl%
\url{https://arxiv.org/abs/2305.07001}
\showURL{%
\tempurl}


\bibitem[Zhang et~al\mbox{.}(2023a)]%
        {zhang2023collm}
\bibfield{author}{\bibinfo{person}{Yang Zhang}, \bibinfo{person}{Fuli Feng}, \bibinfo{person}{Jizhi Zhang}, \bibinfo{person}{Keqin Bao}, \bibinfo{person}{Qifan Wang}, {and} \bibinfo{person}{Xiangnan He}.} \bibinfo{year}{2023}\natexlab{a}.
\newblock \showarticletitle{Collm: Integrating collaborative embeddings into large language models for recommendation}.
\newblock \bibinfo{journal}{\emph{arXiv preprint arXiv:2310.19488}} (\bibinfo{year}{2023}).
\newblock


\bibitem[Zhou et~al\mbox{.}(2020)]%
        {zhou2020improving}
\bibfield{author}{\bibinfo{person}{Kun Zhou}, \bibinfo{person}{Wayne~Xin Zhao}, \bibinfo{person}{Shuqing Bian}, \bibinfo{person}{Yuanhang Zhou}, \bibinfo{person}{Ji-Rong Wen}, {and} \bibinfo{person}{Jingsong Yu}.} \bibinfo{year}{2020}\natexlab{}.
\newblock \showarticletitle{Improving Conversational Recommender Systems via Knowledge Graph based Semantic Fusion}. In \bibinfo{booktitle}{\emph{KDD '20}}.
\newblock


\end{thebibliography}
% \appendix









\end{document}